\documentclass[aps,prb,preprint]{revtex4-1}

\usepackage{amsfonts}
\usepackage{graphicx}
\usepackage{amsmath}

\begin{document}

\title{The equivalence principle and the relative velocity of local inertial frames}

\author{F. Shojai}
\author{A. Shojai}
\affiliation{Department of Physics, University of Tehran, Tehran, Iran\\ \& \\
School of Physics, \\ 
Institute for Research in Fundamental Sciences (IPM), Tehran, Iran} 

\date{\today}

\begin{abstract}
In this paper we explicitly show that in general relativity, the relative velocity of two local inertial frames is always less than the velocity of light. This fact is a by-product of the equivalence principle. The general result is then illustrated within two examples, the FLRW cosmological model and the Schwarzschild metric.
\end{abstract}

\maketitle

\section{Introduction}

Special relativity does not allow faster-than-light velocities. Neither can a particle move faster than light in an inertial frame, nor can the relative velocity of two inertial frames  be greater than that of light. 

Naturally this question arises: what is the situation in general relativity? In Einstein's theory of general relativity, inertial frames are defined locally as freely falling frames. According to the equivalence principle, the special theory of relativity is applicable in each of these local frames. It follows that in these local frames, the velocity of any particle is always less than the velocity of light. However, the situation for the relative velocity of neighboring or far away inertial frames needs more attention. Even for neighboring frames, simply dividing the spatial coordinate distance by the temporal coordinate distance of two frames can result in a superluminal velocity. By spatial and temporal coordinate distance of two neighboring frames we mean the difference between the coordinates of the origin of the two frames in an arbitrary global frame covering both local frames. In special relativity, it is simple to observe that dividing spatial distance by temporal distance can lead to superluminal velocity. 

For example, consider two particles (either near each other or far away) moving with velocity $0.9c$ in different directions. The above mentioned method of calculation of the relative velocity would result in $1.8c$. But using the relativistic velocity addition method (which is equivalent to making a Lorentz transformation to bring one of the particles at rest), one gets $\sim 0.99c$. It is clear that this second method is physically meaningful.

The existence of curvature of the space-time in general relativity, makes the situation more complicated. For example in a spatially flat cosmological model, the spatial distance between two particles at rest at locations $\vec x_1$ and $\vec x_2$ is $d=a(t)|\vec x_2-\vec x_1|$. $a(t)$ is the scale factor of the universe. Differntiating this distance with respect to the time (which is equivalent to the above naive way of defining relative velocity), the Hubble law $v=\dot a d/a$ is obtained. Setting $a\sim t^\alpha$ (which is true both for radiation-dominated universe ($\alpha=1/2$) and matter-dominated universe ($\alpha=2/3$)), the velocity is $v=d/2t$. This is superluminal for $d>2t$. Note that in the early universe ($t\rightarrow 0$), where the curvature is large, the velocity is superluminal even for small $d$. 

The above discussions apply for two particles at rest in this cosmological model. If the two particles move in the opposite directions, the velocity would be more larger than the velocity of light. 

In cosmology, this  is usually explained in this way:  There is no contradiction with special relativity, since  objects do not move in the co-moving frame and only the space expands, and the receding velocity does not correspond to a physical velocity.\cite{line1,line2,line3,cook}  To define the correct relative velocity of two objects in special relativity, one should measure the velocity of the first object in the rest frame of the other object. In order  to define the relative velocity of two frames in general relativity, we have to parallel-transport one of the local frames to the location of the other one. As we shall explain in the next section, by this parallel-transport we mean, transporting all the geometrical objects, like a vector, parallel to themselves by taking into account not only the change in the components of the object, but also the change in the unit vectors of the coordinate frame.

For a spatially flat FLRW cosmological model, it has been  shown that \cite{kaya} the relation between two  freely falling frames is given by a Lorentz transformation in which the relative velocity is less than that of light, both for nearby and far away frames.

In this paper,  we study two nearby or far away local frames. We show that, in general, the parallel-transporting the axes of the first local frame to the location of the second one, and the axes of the second frame are related to each other by a Lorentz transformation. Therefore the relative velocity is always less than that of light. We show that this is  correct not only for FLRW cosmological model as is shown in Ref.~\onlinecite{kaya}, but also for any space-time. Then we shall explicitly evaluate this relative velocity  for two examples, the FLRW and the Schwarzschild metric.

\section{The relation of neighboring local frames}

According to the equivalence principle, it is always possible to find a reference frame, at any point in space-time, such that special relativity is valid at that point and in a sufficiently small neighborhood around it. This frame is called a \textit{freely falling frame}, because no gravity is present in it as it falls freely under the action of gravity. Alternatively, this frame  is called a \textit{local inertial frame}. We use  the notation $\mathbf{e}_a$ for the unit vectors of the coordinate system of this frame. These unit vectors and those of the global coordinate frame are related to each other by \textit{tetrads}, which we shall define later. In choosing this coordinate system one has freedom in two ways.  First, the basis vectors of any local inertial frame can be Lorentz-transformed (not Poincar\'e-transformed, because translations can push the point outside of the neighborhood). Second, it is possible to use any curvilinear local coordinate system, instead of the usual Cartesian one. 

In the absence of gravity, it is possible to choose these local frames such that the unit vectors of the coordinate frames at different points are parallel, and thus special relativity is valid globally. 
Therefore gravity is the nontrivial rotation of the unit vectors of nearby inertial observers. In this sense, gravity is a gauge theory obtained via localizing the global Lorentz group of transformations.\cite{gauge}

\begin{figure}[h!]
\includegraphics[width=10.0cm]{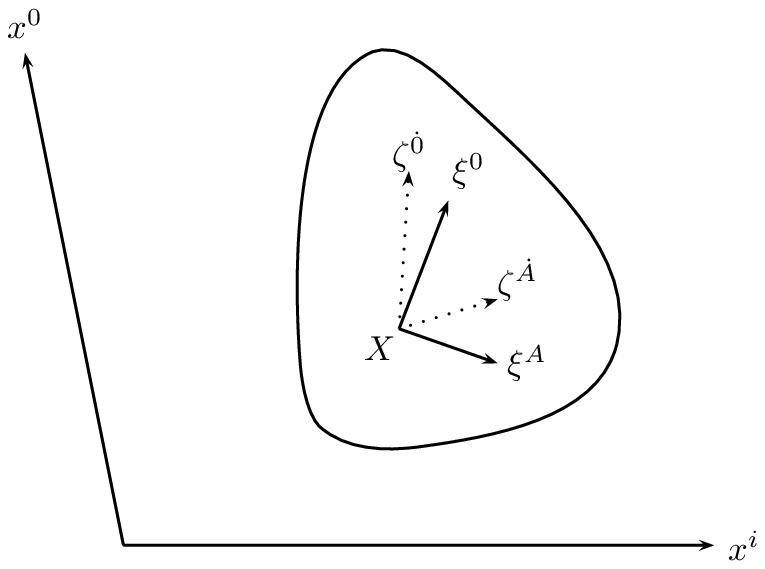}
\caption{Global ($x^\mu$), local Cartesian ($\xi^a$), and local curvilinear ($\zeta^{\dot a}$) coordinate systems near the space-time point~$X$.}
\label{fig1}
\end{figure}

In order to make our notation more clear, let us distinguish between three coordinate systems, as shown in Fig.~\ref{fig1}. 
The global coordinates are denoted by $x^\mu$. The local Cartesian and non-Cartesian coordinates centered around point $X$ are denoted by $\xi_X^a$ and $\zeta_X^{\dot a}$, respectively. We will use Greek indices, $\mu, \nu, \dots$, to represent global coordinates, with 0 the temporal coordinate (in units with the speed of light equal to~1) and the letters $i, j, \ldots$ indicating the spatial coordinates.  For the local Cartesian indices we use $a, b, \dots$ (temporal $0$, spatial $A, B, \dots$), and for local curvilinear coordinates we use $\dot a, \dot b, \dots$ (temporal $\dot 0$, spatial $\dot A, \dot B, \dots$).

For each coordinate system, one can define unit vectors, metric, connection, and lowering and raising indices. These definitions and the relation between them are summarized in Table~\ref{tab1}.

Unit vectors of the global, the local Cartesian, and the local curvilinear frames are denoted by $\mathbf{g}_\mu(x)$, $\mathbf{e}_a(X)$, and $\boldsymbol{\gamma}_{\dot a}(X)$, respectively. Their inner products are defined in the second and third rows of Table~\ref{tab1}.

Since a general displacement vector can be written as $\mathbf{dx}=dx^\mu \mathbf{g}_\mu$, the length of a vector is $\mathbf{dx}\cdot\mathbf{dx}=(\mathbf{g}_\mu\cdot\mathbf{g}_\nu)dx^\mu dx^\nu$, and thus the metric is $g_{\mu\nu}=\mathbf{g}_\mu\cdot\mathbf{g}_\nu$, with signature ($+---$). Similar reasoning leads to the third row of Table~\ref{tab1}.

The relation between global and local Cartesian frames can be obtained using $\mathbf{dx}=dx^\mu \mathbf{g}_\mu=d\xi^a \mathbf{e}_a=(\partial \xi^a/\partial x^\mu)\mathbf{e}_a dx^\mu$. Therefore we can write $\mathbf{g}_\mu=e^a_\mu\mathbf{e}_a$, in which the \textit{tetrads} $e^a_\mu$ are defined as $e^a_\mu=\partial \xi^a/\partial x^\mu$. Similar tetrads relating other local frames are defined in Table~\ref{tab1}.

Finally, the affine connection components are the derivative of unit vectors. It is clear from the fifth row of Table~\ref{tab1} that the global connection consists of two terms. The first term corresponds to the global change of the unit vectors, while the second term is some sort of \textit{internal} connection representing the use of curvilinear coordinates, which is zero if one uses Cartesian coordinates. 

As stated earlier, we have the freedom to make Lorentz transformation on the local coordinates,\begin{equation}
\xi'^a=\Lambda^a_b\xi^b,
\end{equation} 
and the freedom to choose non-Cartesian coordinates:
\begin{equation}
\xi^a\rightarrow \zeta^{\dot a}=\zeta^{\dot a}(\xi).
\end{equation} 
The presence of gravity leads to the rotation of local unit vectors from one point to a neighboring point. But there is a restriction on their relation. We shall show that the  tetrads (and thus the unit vectors) at any two neighboring points are related to each other by a Lorentz transformation. Therefore, the relative velocity of two neighboring freely falling frames is less than the velocity of light. This parallel transportation can be integrated to give the same relation between two far away local frames.

 To show this, let us first parallel-transform a four-vector $\mathbf{S}$ from the point $x$ to a neighboring point $x+dx$. Since $\mathbf{S}=S^\mu\mathbf{g}_\mu$,
\begin{eqnarray}
\delta\mathbf{S} &\equiv& \mathbf{S}(x+dx)-\mathbf{S}(x)=S^\mu(x+dx)\mathbf{g}_\mu(x+dx)-S^\mu(x)\mathbf{g}_\mu(x)\nonumber\\
&=&\left (\mathbf{g}_\mu\partial_\alpha S^\mu+S^\mu\partial_\alpha \mathbf{g}_\mu \right ) dx^\alpha.
\end{eqnarray}
The components of the four-vector (i.e., $\mathbf{g}^\mu\cdot\mathbf{S}$) are thus parallel-transported as
\begin{equation}
S^\mu\longrightarrow \bar S^\mu=
S^\mu(x+dx)+\Gamma^\mu_{\nu\kappa}(x)S^\nu(x)dx^\kappa.
\end{equation}

To obtain the change in the unit vectors, let us set
\begin{equation}
S^\mu\longrightarrow e^\mu_a, \qquad
\bar S^\mu\longrightarrow X^\mu_a.
\end{equation}
Using the relations in Table~\ref{tab1}, we see that the parallel-transported unit vector consists of three terms, the unit vector itself, the term arising from Taylor expanding around the first point and the connection term arising from the change of unit vectors:
\begin{equation}
X^\mu_a=e^\mu_a+dx^\nu\partial_\nu e^\mu_a+\Gamma^\mu_{\nu\kappa} e^\nu_a dx^\kappa .
\end{equation}
Multiplying the above relation by $e_\mu^b(x+dx)$, and retaining terms up to first order in $dx^\mu$, we obtain the same result in terms of local indices:
\begin{equation}
X^b_a=\delta^b_a+dx^\nu e^b_\mu \partial_\nu e^\mu_a +\Gamma^\alpha_{\mu\kappa}e^b_\alpha e^\mu_a dx^\kappa .
\label{e1}
\end{equation}
This relation defines the way that the components of the local unit vectors are parallel-transported to the neighboring points.

The claim that nearby local frames are Lorentz-rotated means that we must have
\begin{equation}
X^b_a=\Lambda^b_{\ a}\delta^c_a=\delta^b_a+\vartheta^b_{\ a},
\label{e2}
\end{equation}
with the Lorentz parameters $\vartheta_{ab}$ being antisymmetric. Combining Eqs.\ (\ref{e1}) and (\ref{e2}), we get
\begin{equation}
\vartheta_{ab}=\Gamma^\mu_{\nu\kappa}e^c_\mu e^\nu_b \eta_{ac} dx^\kappa+\eta_{ac}e^c_\mu dx^\nu \partial_\nu e^\mu_b .
\end{equation}
Assuming metric compatibility, this can be simplified by substituting the metric in terms of tetrads,
\begin{equation}
g_{\mu\nu}=\mathbf{g}_\mu\cdot\mathbf{g}_\nu=e^a_\mu e^b_\nu\eta_{ab},
\end{equation}
in the connection term.
After rearranging the terms and writing $dx^\mu=e^\mu_a d\xi^a$, we get
\begin{equation}
\vartheta_{ab}=\frac{1}{2}d\xi^c\left \{e^\alpha_c\left (\partial_b e_{a\alpha}-\partial_a e_{b\alpha}\right)+ 
 \left ( e^\alpha_a\partial_b e_{c\alpha}-e^\alpha_b\partial_a e_{c\alpha}  \right ) + \left ( e_{a\alpha}\partial_c e^\alpha_b - e_{b\alpha}\partial_c e^\alpha_a\right )\right \},
\label{main}
\end{equation}
which is clearly antisymmetric.

Therefore the conclusion is that for any general space-time, any two nearby freely falling observers are Lorentz transformed with respect to each other and thus their relative velocity is less than that of light. 
To obtain the relative velocity of two far away local frames, one should integrate the above result. That is the relation between two far away local frames is given by multiplication of a large number of Lorentz transformations which we know from special relativity that is a Lorentz transformation. This means that the relative velocity of two far away local frames is subluminal, too. 

In the next section, we shall illustrate this general result within two examples, FLRW cosmological model and Schwarzschild solution.

\section{Examples}

To illustrate the result of the previous section, let us  apply it to some examples and see how subluminal velocities are obtained.

\subsection{FLRW cosmology}

As a first example, suppose that one wants to calculate the relative velocity of two galaxies in the Friedmann-Lema\^itre-Robertson-Walker geometry. Since  we are using the co-moving frame in cosmology, the relative velocity of two galaxies is just the relative velocity of local inertial frames.   
Let us first parallel-transport the four-velocity $v^\nu$ at an arbitrary point to a neighboring point along a curve whose space-like tangent vector is $ k^\mu=(0,\vec k$). We have
\begin{equation}
k^\mu\partial_\mu v^\nu+\Gamma^\nu_{\mu\alpha}k^\mu v^\alpha=0 .
 \end{equation}
Using the general form of an expanding universe space-time metric,
 \begin{equation}
 g_{\mu\nu}=
 \begin{pmatrix}
	1 & 0\\
	0 & -a^2(t) h_{ij}\\
\end{pmatrix},
\end{equation}
in which $h_{ij}$ is the three-metric of spatial slices, 
and noting that the only non-vanishing components of the connection are
\begin{equation}
\Gamma^0_{ij}=a\dot{a}h_{ij},\qquad  \Gamma^i_{0j}=\Gamma^i_{j0}=\frac{\dot{a}}{a}\delta^i_j,\qquad \Gamma^i_{jk}=\gamma^i_{jk},
\end{equation}
where $\gamma^i_{jk}$ are the three-connection of the three-metric $h_{ij}$, the spatial part of the equation of parallel transportation is:
\begin{equation}
k^i\partial_i v^0+k^i v^j h_{ij} a \dot{a}=0
\end{equation}
while the temporal part reads as:
\begin{equation}
k^i\partial_i v^j+k^j v^0 \frac{\dot{a}}{a}+k^i v^k \gamma^j_{ik}=0 .  
\end{equation}

Using the relation $g_{\mu\nu}=e^a_\mu e^b_\nu \eta_{ab}$, the tetrads are given by
\begin{equation}
e^0_\mu=(1,\vec 0)
\end{equation}
and
\begin{equation}
e^I_\mu=a(0, E^I_i),
\end{equation}
where $E^I_i$ are the triads of the metric $h_{ij}$, defined as
\begin{equation}
E^I_iE^J_j\delta_{IJ}=h_{ij} .
\end{equation}
The velocity components in the local inertial frames defined as $u^A=e^A_\mu v^\mu$ are given by
\begin{equation}
u^0=e^0_\mu v^\mu=v^0 ,
\end{equation}
\begin{equation}
u^I=e^I_\mu v^\mu=a E^I_i v^i .
\end{equation}
Then, the parallel-transport equations of velocity along $k^A$ are given by:
\begin{eqnarray}
\frac{\partial u^0}{\partial\ell}+\frac{\dot{a}}{a} u&=&0 ,\\
\frac{\partial u}{\partial\ell}+\frac{\dot{a}}{a} u^0&=&0 ,
\end{eqnarray}
in which $u=k_I u^I$ and $\partial/\partial\ell= k^i\partial/\partial x^i= k^I\partial/\partial \xi^I$, and $\ell$ is the local distance of two nearby galaxies. 

The above coupled equations can be solved easily to obtain
\begin{equation}
 \begin{pmatrix}
	u^0\\ 
	u\\
\end{pmatrix}
=\exp \left [ -
\begin{pmatrix}
	0 & 1\\ 
	1 & 0\\
\end{pmatrix}
\frac{\dot{a}}{a} \ell \right ]
 \begin{pmatrix}
	1\\ 
	0\\
\end{pmatrix} \text{.}
\end{equation}
Diagonalizing the exponential leads to
\begin{equation}
 \begin{pmatrix}
	u^0\\ 
	u\\
\end{pmatrix}
= 
\begin{pmatrix}
	\cosh(\dot{a}\ell/a) & -\sinh(\dot{a}\ell/a)\\ 
	-\sinh(\dot{a}\ell/a) & \cosh(\dot{a}\ell/a)\\
\end{pmatrix}
 \begin{pmatrix}
	1\\ 
	0\\
\end{pmatrix} .
\end{equation}
 
This expression shows that the transported velocity is boosted with respect to the local inertial frame, with the relative velocity
\begin{equation}
v_\text{rel}=\tanh \Bigl(\frac{\dot{a}}{a}\ell\Bigr),
\label{rel}
\end{equation}
which is less than the velocity of light. This result extends the derivation of Ref.~\onlinecite{kaya} to a geometry with arbitrary spatial curvature.

This result can also be obtained from the general calculations of the previous section. Simply substituting the tetrads of the FLRW metric in equation (\ref{main}), one gets
\begin{equation}
\vartheta_{0I}=\frac{\dot{a}}{a}d\xi_I
\end{equation}
and 
\begin{equation}
\vartheta_{IJ}=0.
\end{equation}
This last equation which shows the vanishing of the space-space components of the Lorentz parameters, indicates that the nearby local frames are not spatially rotated with respect to each other. In addition, keeping in mind that the relative velocity and the boost parameter are related to each other via $v_{rel}=\tanh \vartheta_{0I}$, we see that the relative velocity of two nearby local frames is less than that of light.  
To obtain the relative velocity of two distant local frames (e.g. the relative velocity of two distant galaxies), one should make successive Lorentz transformations which leads to the boost parameter $\vartheta_{0I}=\int \dot{a}d\xi_I/{a}=\dot{a}\ell/{a}$. This gives the subluminal relative velocity given by Eq.~(\ref{rel}).

It has to be noted that, if we calculate the velocity by the naive way of dividing the spatial distance by temporal distance, we get the velocity given by Hubble's law $\dot a \ell/a$, as discussed in the Introduction.  Therefore, the physical relative velocity given by Eq.~(\ref{rel}) is always less than the velocity of light, while the naive calculation can lead to superluminal velocities. 

\subsection{Schwarzschild black hole}

As a second example, let us study the Schwarzschild metric:
 \begin{equation}
 g_{\mu\nu}=
 \begin{pmatrix}
	A(r) & 0 & 0 & 0\\
	0 & -1/A(r) & 0 & 0\\
	0 & 0 & -r^2 & 0 \\
	0 & 0 & 0 & -r^2 \sin^2\theta \\
\end{pmatrix},
\end{equation} 
with $A(r)=1-r_s/r$, in which $r_s=2GM$ is the Schwarzschild radius. The tetrads are
\begin{equation}
e^0_\mu=(\sqrt{A},0,0,0),\quad 
e^1_\mu=(0,\frac{1}{\sqrt{A}},0,0),\quad
e^2_\mu=(0,0,r,0),\quad
e^3_\mu=(0,0,0,r\sin\theta),
\end{equation}
and the non-vanishing components of the connection are
\begin{equation}
\begin{gathered}
\Gamma^0_{10}=\frac{A'}{2A},\quad
\Gamma^1_{00}=\frac{1}{2}AA',\quad
\Gamma^1_{11}=-\frac{A'}{2A},\quad
\Gamma^1_{22}=-rA,
\\
\Gamma^1_{33}=-r\sin^2\theta A,\quad
\Gamma^2_{12}=\frac{1}{r},\quad 
\Gamma^2_{33}=-\sin\theta\cos\theta.
\end{gathered}
\end{equation}
Using the Eq.~(\ref{main}), the only nonzero boost component of two nearby freely falling frames is
\begin{equation}
\vartheta_{01}=\frac{A'}{2\sqrt{A}}d\xi^0=\frac{r_s}{2r^2\sqrt{1-r_s/r}}d\xi^0=\frac{r_s}{2r^2}dx^0,
\label{xyz}
\end{equation}
leading to the relative velocity
\begin{equation}
v_\textrm{rel}=\tanh\left(\frac{A'\xi^0}{2\sqrt{A}}\right),
\label{xyz1}
\end{equation}
which is again less than that of light.

Like the previous example, we can obtain the same result from the parallel-transport equations. Choosing $k^\mu=(1,0,0,0)$ in the parallel-transport relation and going to the local inertial frame, one gets the following relations:
\begin{equation}
\frac{\partial u^0}{\partial \xi^0}+\frac{A'}{2\sqrt{A}} u^1=0,
\end{equation}
\begin{equation}
\frac{\partial u^1}{\partial \xi^0}+\frac{A'}{2\sqrt{A}} u^0=0.
\end{equation}
The solution is
\begin{equation}
 \begin{pmatrix}
	u^0\\ 
	u^1\\
\end{pmatrix}
=\exp \left [ -
\begin{pmatrix}
	0 & 1\\ 
	1 & 0\\
\end{pmatrix}
\frac{A'}{2\sqrt{A}}\xi^0 \right ]
 \begin{pmatrix}
	1\\ 
	0\\
\end{pmatrix},
\end{equation} 
which can be evaluated via diagonalization to obtain
\begin{equation}
 \begin{pmatrix}
	u^0\\ 
	u\\
\end{pmatrix}
= 
\begin{pmatrix}
	\cosh(A'\xi^0/2\sqrt{A}) & -\sinh(A'\xi^0/2\sqrt{A})\\ 
	-\sinh(A'\xi^0/2\sqrt{A}) & \cosh(A'\xi^0/2\sqrt{A})\\
\end{pmatrix}
 \begin{pmatrix}
	1\\ 
	0\\
\end{pmatrix}.
\end{equation}
This is a Lorentz transformation with the relative velocity given by Eq.~(\ref{xyz1}).

\section{Conclusions}

According to the equivalence principle, the special theory of relativity holds in local freely falling frames.
It is an important property of general relativity that these local inertial frames are related to each other by Lorentz transformations. This means that the relative velocity of local frames is always less than the velocity of light.
  
The naive way of obtaining the velocity, i.e. dividing the spatial distance by the temporal one, can lead to superluminal velocity. But this is not true for a physical relative velocity which is obtained by parallel transporting the first local frame to the location of the second one and comparing them. The physical relative velocity is always less than the velocity of light.

\begin{acknowledgments}
This work was supported by a grant from University of Tehran, and also by a grant from the Institute for Research in Fundamental Sciences (IPM).
\end{acknowledgments}

\newpage
\section*{Tables}

\begin{table}[h!]
\caption{Geometrical relations for different frames.}
\vspace{6 pt}
\begin{tabular}{lccc}
\hline\hline
	& Global & Local Cartesian & Local curvilinear \\
	& frame & frame & frame \\
\hline
	Coordinates & $x^\mu$ & $\xi^a$ & $\zeta^{\dot a}$ \\
	Unit vectors & $\mathbf{g}_\mu(x)$ & $\mathbf{e}_a(X)$ & $\boldsymbol{\gamma}_{\dot a}(X)$ \\

	 & $\mathbf{g}^\mu\cdot \mathbf{g}_\nu=\delta^\mu_\nu$ & $\mathbf{e}^a\cdot \mathbf{e}_b=\delta^a_b$ & $\boldsymbol{\gamma}^{\dot a}\cdot \boldsymbol{\gamma}_{\dot b}=\delta^{\dot a}_{\dot b}$ \\
	Metric & $\mathbf{g}_\mu\cdot \mathbf{g}_\nu=g_{\mu\nu}$ & $\mathbf{e}_a\cdot \mathbf{e}_b=\eta_{ab}$ & $\boldsymbol{\gamma}_{\dot a}\cdot \boldsymbol{\gamma}_{\dot b}=\gamma_{{\dot a}{\dot b}}$ \\
	Change of frame & $\mathbf{g}_\mu=e^a_\mu \mathbf{e}_a=\omega^{\dot a}_\mu\boldsymbol{\gamma}_{\dot a}$ & $\mathbf{e}_a=e^\mu_a \mathbf{g}_\mu=\epsilon^{\dot b}_a\boldsymbol{\gamma}_{\dot b}$ & $\boldsymbol{\gamma}_{\dot a}=\epsilon ^b_{\dot a} \mathbf{e}_b=\lambda^\mu_{\dot a}\mathbf{g}_\mu$ \\
	Tetrads & $e^a_\mu\equiv \frac{\partial \xi^a}{\partial x^\mu}$ & $e^\mu_a\equiv \frac{\partial x^\mu}{\partial \xi^a}$ & $\epsilon^b_{\dot a}\equiv \frac{\partial \xi^b}{\partial \zeta^{\dot a}}=e^b_\mu\omega^\mu_{\dot a}$ \\
	Inverses & $\omega^{\dot a}_\mu\equiv \frac{\partial \zeta^{\dot a}}{\partial x^\mu}$ & $\epsilon^{\dot b}_a\equiv \frac{\partial \zeta^{\dot b}}{\partial \xi^a}=\omega^{\dot b}_\mu e^\mu_a$ & $\lambda^\mu_{\dot a}\equiv \frac{\partial x^\mu}{\partial \zeta^{\dot a}}=e^\mu_b\epsilon^b_{\dot a}$ \\
	Connection & $\Gamma^\alpha_{\mu\nu}=\mathbf{g}^\alpha\cdot\frac{\partial\mathbf{g}_\nu}{\partial x^\mu}$ & $0$ & $\dot{\Gamma}^{\dot a}_{\dot b \dot c}=\boldsymbol{\gamma}^{\dot a}\cdot \frac{\partial \boldsymbol{\gamma}_{\dot c}}{\partial\zeta^{\dot b}}$ \\
	 & $= e^\alpha_a\frac{\partial e^a_\nu}{\partial x^\mu}= \gamma^\alpha_{\dot a}\frac{\partial \gamma^{\dot a}_\nu}{\partial x^\mu}+$ &  & $= \epsilon^{\dot a}_a\frac{\partial \epsilon^a_{\dot c}}{\partial \zeta^{\dot b}}$ \\
	 & $\gamma^\alpha_{\dot a}\gamma^{\dot b}_\mu\gamma^{\dot c}_\nu\dot{\Gamma}^{\dot a}_{\dot b \dot c}$ &  &  \\
	Metric compatibility & $\Gamma^\alpha_{\mu\nu}=\frac{1}{2}g^{\alpha\beta}\times$ & & $\dot{\Gamma}^{\dot a}_{\dot b\dot c}=\frac{1}{2}\gamma^{\dot a\dot d}\times$ \\
	 & $\left ( \frac{\partial g_{\mu\beta}}{\partial x^\nu}+\frac{\partial g_{\nu\beta}}{\partial x^\mu} - \frac{\partial g_{\mu\nu}}{\partial x^\beta}\right )$ & & $\left ( \frac{\partial \gamma_{\dot b\dot d}}{\partial \zeta^{\dot c}}+\frac{\partial \gamma_{\dot c\dot d}}{\partial \zeta^{\dot b}} - \frac{\partial \gamma_{\dot b\dot c}}{\partial \zeta^{\dot d}}\right )$\\
	\hline\hline
\end{tabular}
\label{tab1}
\end{table}

\newpage
\section*{Figure Captions}

\begin{thebibliography}{99}
\bibitem{line1}
C.~H.~Lineweaver and T.~M.~Davis, ``Misconceptions about the big bang,'' Sci. Am. \textbf{292}(3), 36--45 (2005).
\bibitem{line2}
T.~M.~Davis and C.~H.~Lineweaver, ``Expanding Confusion: Common Misconceptions of Cosmological
Horizons and the Superluminal Expansion of the Universe,'' Publ. Astron. Soc. Aust. \textbf{21}, 97--109 (2004).
\bibitem{line3}
T.~M.~Davis, C.~H.~Lineweaver, and J.~K.~Webb, ``Solutions to the tethered galaxy problem in an expanding universe and the observation of receding blueshifted objects,'' Am. J. Phys. \textbf{71}, 358--364 (2003).
\bibitem{cook}
R. Cook and M. S. Burns, ``Interpretation of the cosmological metric,'' Am. J. Phys. \textbf{77}, 59--66 (2009).
\bibitem{kaya}
A. Kaya, ``Hubble's law and faster than light expansion speeds,'' Am. J. Phys. \textbf{79}, 1151--1154 (2011).
\bibitem{gauge}
See, e.g.,  M.~Blagojevic, \textit{Gravitation and Gauge Symmetries} (CRC Press, 2001), and M.~Blagojevic and F.~W.~Hehl, editors, \textit{Gauge Theories of Gravitation: A Reader with Commentaries} (Imperial College Press, London, 2013).
\end{thebibliography}
\end{document}